\documentclass[prd,reprint,showpacs,showkeys]{revtex4}
\usepackage{graphics}
\usepackage{eurosym}
\usepackage{amsfonts}
\usepackage{mdframed}
\usepackage{amssymb}
\usepackage{amsmath}
\usepackage{graphicx}
\usepackage[font={footnotesize,it}]{caption}
\usepackage{hyperref}

\begin{document}

\title{Massive Vector Particles Tunneling From Noncommutative Charged Black
Holes and its GUP-corrected Thermodynamics}
\author{Ali \"{O}vg\"{u}n}
\email{ali.ovgun@emu.edu.tr}
\affiliation{Physics Department, Eastern Mediterranean University, Famagusta, Northern
Cyprus, via Mersin 10, Turkey}
\author{Kimet Jusufi}
\email{kimet.jusufi@unite.edu.mk}
\affiliation{Physics Department, State University of Tetovo, Ilinden Street nn, 1200,
Macedonia}
\date{\today }

\begin{abstract}
In this paper, we investigate the tunneling process of charged massive
bosons $W^{\pm }$ (spin-1 particles) from noncommutative charged black holes
such as charged RN black holes and charged BTZ black holes. By applying the
WKB approximation and by using the Hamilton-Jacobi equation we derive the
tunneling rate and the corresponding Hawking temperature for those black
holes configuration. Furthermore, we show the quantum gravity effects using the GUP on the Hawking temperature for the  noncommutative RN  black holes. The tunneling rate shows that the radiation deviates
from pure thermality and is consistent with an underlying unitary theory. 


\end{abstract}
\maketitle
\section{Introduction}

Hawking's amazing discovery that black holes should radiate thermally,
undoubtedly, remains as one of the most interesting results in physics.
Basically, Hawking used semiclassical arguments and showed that black holes
must radiate thermally just as a black body \cite{Hawking1}. Furthermore,
due to the thermal nature of this radiation the initial pure state is mapped
to a final mixed state in a non-unitary process which leads to the famous
information loss paradox.

Over the years, people were able to derive the Hawking temperature by
applying different techniques and methods \cite{gibbons}. Among others, the
quantum tunneling has attracted a great interest. The so called null
geodesic approach, introduced and developed by Kraus-Wilczek-Parkih \cite%
{sh1,kraus1,kraus2,perkih1,perkih2,perkih3} and the Hamilton-Jacobi approach
proposed and developed by Angheben-Padmanabhan \cite{ang,sri1,sri2}. A great
number of different black hole configurations have been studied in details
showing that the Hawking temperature does not depend on the particular
coordinate transformation we use and the nature of the particles being
emitted from black holes. In particular, the tunneling of scalar particles 
\cite{ahmed,kimet}, spin-1/2 particles \cite{vanzo,mann0,mann1,mann2},
spin-3/2 \cite{mann3,sh2} and spin-1 particles \cite%
{kruglov1,kruglov2,sh11,sh3,sh4,huang,chenn}, have been successfully
investigated. Furthermore, recently quantum gravity effects of the entangled
particles tunneling is investigated by using the generalized uncertainty
principle \cite{ao}. The calculations shows that the tunneling rate deviates
from pure thermality and is consistent with unitary theory.

\textit{On the other hand, since the discovery of Banados-Teitelboim-Zanelli
(BTZ) black hole\cite{btz}, three dimensional black holes become popular
research area of quantum gravity. Today, the three dimensional gravity is
used to test the theories \ behind the AdS/CFT correspondence. In Ref. \cite%
{brito}, the authors analyzed the gravitational Aharonov-Bohm effect due to
BTZ black hole in a noncommutative background. Moreover, in Ref. \cite{jafer}%
, has been analyzed the behavior of a particle test in the noncommutative
BTZ space-time. The thermodynamic properties of the charged BTZ black hole
were investigated in Ref. \cite{Hendi}.}

Recently, noncommutative gravity has attracted a lot of interest. Several
black hole solutions have been found by using a new model of commutativity,
including the neutral and charged black holes \cite{nicolini1,nicolini2},
inspired higher-dimensional charged black holes \cite{nicolini3}, and
rotating charged black holes \cite{nicolini4}. Furthermore, it is shown that
this model of noncommutativity preserves the Lorentz invariance, Unitarity
and UV-finiteness of quantum field theory \cite{sma1,sma2,sma3}. According
to this scenario, the black hole evaporation process might finish when the
black hole reaches some minimal mass $M_{0}$, this on the other hand might
shed some light on the information loss paradox. The key idea in
noncommutative gravity is the noncommutativity of the spacetime points. This
noncomutativity can be encoded in terms of noncommutative self-adjoint
operators given by the commutator%
\begin{equation}
\lbrack x^{\mu },x^{\nu }]=i\theta ^{\mu \nu }
\end{equation}%
where $\theta ^{\mu \nu }$ is an anti-symmetric second-rank tensor. One
natural consequence of this noncommutativity is the minimum length in
spacetime. It should be noted that the point matter/charge distribution
should be expressed by a Gaussian function rather than a Dirac delta
function. The quantum tunneling of Hawking radiation also has been
investigated in context of noncommutative black holes \cite%
{hamid,zhao,nozari1,nozari2}.\textit{ As we know from particle physics, a massive
boson is a spin-1 particle described by the Proca equation. In particular,
the charged massive bosons i.e., $W^{\pm }$ and Z are very important in the
Standard Model of particle physics, they are known as a force carriers of
the weak interaction and for the important role in the confirmed Higgs
Boson. In this context it's interesting to note the possible existence of a
massive photon or the so-called Darklight \cite{kahn,bednyakov}, and the
charged spin-1 boson speculated as a potential dark matter candidates \cite%
{balewski,emidio}. On the other hand, due to the nontrivial interaction
between the charged massive boson ($W^{\pm }$) fields and the
electromagnetic field may lead to new insights. Therefore, we think the
study of spin-1 particles merits more attention, in particular, in the
context of black hole physics.}

\textit{During the last few years, modification of the uncertainty principle
to a generalized uncertainty principle (GUP) becomes important tool to
investigate \ the effect of the quantum gravity on many different research
areas\cite{gup1,gup2,gup3,gup4}. Divergences in physics is suggested that it
can be solved easily by using GUP relation instead of the using of usual
Heisenberg uncertainty relation. Recently in Ref.(\cite{brit}), the authors
study the corrections for the entropy of the noncommutative BTZ\ black
holes. In this paper, using the same procedure we investigate the
GUP-corrected thermodynamics of charged black holes. }

\textit{Inspired by what has just been said, in this work we follow \cite{xiang},
and study the tunneling of charged massive bosons ($W^{\pm }$) from
noncommutative charged black holes. The field equations can be derived by
using the Lagrangian given by the Glasgow-Weinberg-Salam model. We then use
the WKB approximation and the separation of variables which results with a
set of four linear equations. Solving for the radial part by using the
determinant of the metric equals zero, we found the tunneling rate and the
corresponding Hawking temperature. Finally, we aim to compute the corrected
Hawking temperature and entropy by incorporating the noncommutative
parameter $\theta $ and quantum gravity effects $\alpha $.}

The paper is organized as follows. In Sec. II, we investigate the tunneling
of massive vector particles from the noncommutative charged Reissner-Nordstr%
\"{o}m (RN) black hole and calculate the corresponding Hawking radiation. In
Sec. III, we extend the tunneling of massive vector particles from
noncommutative BTZ black holes and compute the corresponding Hawking
temperature. In Sec. IV, we study the GUP-corrected Hawking temperature then
we compute the GUP-Entropy corrections for the noncommutative charged RN
black hole. Finally, in Sec. V, we comment on our results.

\section{Tunneling From Noncommutative RN Black Holes}

In noncommutative gravity the mass and charge distributions cannot be
localized as point like particles but rather diffused and therefore given by
a Gaussian distribution of some minimal width $\sqrt{\theta }$, as follows 
\begin{equation}
\rho _{matt.}(r)=\frac{M}{\left( 4\pi \theta \right) ^{3/2}}e^{-\frac{r^{2}}{%
4\theta }}  \label{matd}
\end{equation}%
respectively, 
\begin{equation}
\rho _{el.}(r)=\frac{Q}{\left( 4\pi \theta \right) ^{3/2}}e^{-\frac{r^{2}}{%
4\theta }}.  \label{chad}
\end{equation}

Here, the parameter $\theta $ encodes the minimal length due to the
noncommutativity of the spacetime points. The line element which solves the
Einstein's field equation in noncommutative spacetime is given by \cite%
{nicolini2} 
\begin{equation}
ds^{2}=-f(r)dt^{2}+f(r)^{-1}dr^{2}+r^{2}\left( d\theta ^{2}+\sin ^{2}\theta
d\varphi ^{2}\right) ,  \label{metric1}
\end{equation}%
where 
\begin{equation}
f(r)=\left( 1-\frac{2M_{\theta }}{r}+\frac{Q_{\theta }^{2}}{r^{2}}\right) .
\label{f}
\end{equation}

Using \eqref{matd} and \eqref{chad}, the smeared mass and charge
distributions can be expressed in terms of the lower incomplete Gamma
functions as follows 
\begin{eqnarray}
M_{\theta } &=&\frac{2M}{\sqrt{\pi }}\gamma \left( \frac{3}{2},\frac{r^{2}}{%
4\theta }\right) ,  \label{mm} \\
Q_{\theta } &=&\frac{Q}{\sqrt{\pi }}\sqrt{\gamma ^{2}\left( \frac{1}{2},%
\frac{r^{2}}{4\theta }\right) -\frac{r}{\sqrt{2\theta }}\gamma \left( \frac{1%
}{2},\frac{r^{2}}{2\theta }\right) +r\sqrt{\frac{2}{\theta }}\gamma \left( 
\frac{3}{2},\frac{r^{2}}{4\theta }\right) }, \\
\gamma \left( \frac{a}{b},u\right)  &=&\int_{0}^{x}\frac{du}{u}u^{a/b}e^{-u}.
\end{eqnarray}

Clearly, in the limit $\theta\to 0$, the noncommutative mass and charge
solution reduces to the commutative case $M_{\theta}\to M$ and similarly $%
Q_{\theta}\to Q$. Solving for $f(r=r_{\theta})=0$, we end up with the outer
and inner horizons given by 
\begin{equation}
r_{\theta\pm}=M_{\theta \pm}\pm \sqrt{M^{2}_{\theta \pm}-Q^{2}_{\theta\pm}}.
\end{equation}

Now, recall that the Lagrangian density, which describes the $W$-bosons in a
background electromagnetic field can be written as \cite{xiang} 
\begin{equation}
\mathcal{L}=-\frac{1}{2}\left( D_{\mu }^{+}W_{\nu }^{+}-D_{\nu }^{+}W_{\mu
}^{+}\right) \left( D^{-\mu }W^{-\nu }-D^{-\nu }W^{-\mu }\right) +\frac{%
m_{W}^{2}}{\hbar ^{2}}W_{\mu }^{+}W^{-\mu }-\frac{i}{\hbar }eF^{\mu \nu
}W_{\mu }^{+}W_{\nu }^{-},
\end{equation}%
where $D_{\pm \mu }=\nabla _{\mu }\pm \frac{i}{\hbar }eA_{\mu }$. Here $%
\nabla _{\mu }$ is the covariant derivative, $e$ gives the charge of the $%
W^{+}$ boson, and $A_{\mu }$ is the electromagnetic potential of the black
hole with components $A_{\mu }=(A_{0},0,0,0)$. Using the above Lagrangian
the equation of motion for the $W$-boson field reads 
\begin{equation}
\frac{1}{\sqrt{-g}}\partial _{\mu }\left[ \sqrt{-g}\left( D^{\pm \nu }W^{\pm
\mu }-D^{\pm \mu }W^{\pm \nu }\right) \right] \pm \frac{ieA_{\mu }}{\hbar }%
\left( D^{\pm \nu }W^{\pm \mu }-D^{\pm \mu }W^{\pm \nu }\right) +\frac{%
m_{W}^{2}}{\hbar ^{2}}W^{\pm \nu }\pm \frac{i}{\hbar }eF^{\nu \mu }W_{\mu
}^{\pm }=0  \label{proca1}
\end{equation}%
where $F^{\mu \nu }=\nabla ^{\mu }A^{\nu }-\nabla ^{\nu }A^{\mu }$. In this
paper we will study the tunneling of $W^{+}$ boson, therefore we need to
solve the following equation 
\begin{equation}
\frac{1}{\sqrt{-g}}\partial _{\mu }\left[ \sqrt{-g}g^{\mu \beta }g^{\nu
\alpha }\left( \partial _{\alpha }W_{\beta }^{+}-\partial _{\beta }W_{\alpha
}^{+}+\frac{i}{\hbar }eA_{\alpha }W_{\beta }^{+}-\frac{i}{\hbar }eA_{\beta
}W_{\alpha }^{+}\right) \right]  \label{proca2}
\end{equation}

\begin{equation*}
+\frac{ieA_{\mu }g^{\mu \beta }g^{\nu \alpha }}{\hbar }\left( \partial
_{\alpha }W_{\beta }^{+}-\partial _{\beta }W_{\alpha }^{+}+\frac{i}{\hbar }%
eA_{\alpha }W_{\beta }^{+}-\frac{i}{\hbar }eA_{\beta }W_{\alpha }^{+}\right)
+\frac{m_{W}^{2}g^{\nu \beta }}{\hbar ^{2}}W_{\beta }^{+}+\frac{i}{\hbar }%
eF^{\nu \alpha }W_{\alpha }^{+}=0
\end{equation*}%
for $\nu =0,1,2,3$. Using the WKB approximation 
\begin{equation}
W_{\mu }^{+}(t,r,\theta ,\varphi )=C_{\mu }(t,r,\theta,\varphi)\exp \left( 
\frac{i}{\hbar }S(t,r,\theta ,\varphi )\right) ,  \label{ans1}
\end{equation}%
and taking the lowest order of $\hbar $, we end up with a set of four
equations: 
\begin{eqnarray}
0 &=&C_{0}\left( -(\partial _{1}S)^{2}-\frac{(\partial _{2}S)^{2}}{r^{2}f}-%
\frac{(\partial _{3}S)^{2}}{r^{2}f\sin ^{2}\theta }-\frac{m^{2}}{f}\right)
+C_{1}\left( (\partial _{1}S)\left( eA_{0}+\partial _{0}S\right) \right)
+C_{2}\left( \frac{(\partial _{2}S)}{r^{2}f}\left( \partial
_{0}S+eA_{0}\right) \right)  \notag \\
&+&C_{3}\left( \frac{(\partial _{3}S)}{r^{2}f\sin ^{2}\theta }\left(
\partial _{0}S+eA_{0}\right) \right) ,
\end{eqnarray}%
\begin{eqnarray}
0 &=&C_{0}\left( -(\partial _{1}S)(eA_{0}+\partial _{0}S)\right)
+C_{1}\left( -f\frac{(\partial _{2}S)^{2}}{r^{2}}-f\frac{(\partial _{3}S)^{2}%
}{r^{2}\sin ^{2}\theta }+(\partial _{0}S+eA_{0})^{2}-m^{2}f\right)
+C_{2}\left( f\frac{\partial _{1}S\partial _{2}S}{r^{2}}\right)  \notag \\
&+&C_{3}\left( f\frac{\partial _{1}S\partial _{3}S}{r^{2}\sin ^{2}\theta }%
\right) ,
\end{eqnarray}%
\begin{eqnarray}
0 &=&C_{0}\left( -\partial _{2}S\left( \frac{\partial _{0}S+eA_{0}}{f}%
\right) \right) +C_{1}\left( f(\partial _{2}S)(\partial _{1}S)\right)
+C_{2}\left( -f(\partial _{1}S)^{2}-\frac{(\partial _{3}S)^{2}}{r^{2}\sin
^{2}\theta }+\frac{(\partial _{0}S+eA_{0})^{2}}{f}-m^{2}\right)  \notag \\
&+&C_{3}\left( \frac{(\partial _{2}S)(\partial _{3}S)}{r^{2}\sin ^{2}\theta }%
\right) ,
\end{eqnarray}%
\begin{eqnarray}
0 &=&C_{0}\left( -\partial _{3}S\left( \frac{\partial _{0}S+eA_{0}}{f}%
\right) \right) +C_{1}\left( f(\partial _{3}S)(\partial _{1}S)\right)
+C_{3}\left( -f(\partial _{1}S)^{2}-\frac{(\partial _{2}S)^{2}}{r^{2}}+\frac{%
(\partial _{0}S+eA_{0})^{2}}{f}-m^{2}\right)  \notag \\
&+&C_{2}\left( \frac{\partial _{2}S\partial _{3}S}{r^{2}}\right) .
\end{eqnarray}
Due to spherical symmetry, one can choose $\theta =\pi /2$, the non-zero
elements of the coefficient matrix $\Xi $ are given by 
\begin{eqnarray}
\Xi _{11} &=&\left(-(\partial _{1}S)^{2}-\frac{(\partial _{2}S)^{2}}{r^{2}f}-%
\frac{(\partial _{3}S)^{2}}{r^{2}f}-\frac{m^{2}}{f}\right) \notag \\
\Xi _{12} &=&-\Xi _{21}=(\partial _{1}S)\left( eA_{0}+\partial _{0}S\right) 
\notag \\
\Xi _{13} &=&\left( \frac{(\partial _{2}S)}{r^{2}f}\left( \partial
_{0}S+eA_{0}\right) \right)  \notag \\
\Xi _{14} &=&\left( \frac{(\partial _{3}S)}{r^{2}f}\left( \partial
_{0}S+eA_{0}\right) \right)  \notag \\
\Xi _{22} &=&\left(-f\frac{(\partial _{2}S)^{2}}{r^{2}}-f\frac{(\partial
_{3}S)^{2}}{r^{2}}+(\partial _{0}S+eA_{0})^{2}-m^{2}f\right)  \notag \\
\Xi _{23} &=&f\frac{\left( \partial _{1}S\right) \left( \partial _{2}S\right)%
}{r^{2}}  \notag \\
\Xi _{24} &=&f\frac{\left( \partial _{1}S\right) \left( \partial _{3}S\right)%
}{r^{2}}  \notag \\
\Xi _{31} &=&-\frac{\left( \partial _{0}S+eA_{0}\right) \partial _{2}S}{f} 
\notag \\
\Xi _{32} &=&f(\partial _{2}S)(\partial _{1}S)  \notag \\
\Xi _{33} &=&\left( -f(\partial _{1}S)^{2}-\frac{(\partial _{3}S)^{2}}{r^{2}}%
+\frac{(\partial _{0}S+eA_{0})^{2}}{f}-m^{2}\right)  \notag \\
\Xi _{34} &=&\Xi _{43}=\frac{(\partial _{2}S)(\partial _{3}S)}{r^{2}}  \notag
\\
\Xi _{41} &=&-\frac{\partial _{3}S\left( \partial _{0}S+eA_{0}\right) }{f} 
\notag \\
\Xi _{42} &=&f(\partial _{3}S)(\partial _{1}S)  \notag \\
\Xi _{44} &=&\left( -f(\partial _{1}S)^{2}-\frac{(\partial _{2}S)^{2}}{r^{2}}%
+\frac{(\partial _{0}S+eA_{0})^{2}}{f}-m^{2}\right) .
\end{eqnarray}

Using the space-time symmetries of the metric \eqref{metric1}, we can now
choose the following ansatz 
\begin{equation}
S=-Et+W(r)+j\varphi +H(\theta )+C,
\end{equation}%
and the nontrivial solution of this equation \cite{kruglov1} 
\begin{equation}
\Xi (C_{0},C_{1},C_{2},C_{3})^{T}=0,  \label{matrixeq}
\end{equation}

is obtained by using the the determinant of the matrix $\Xi $ equals zero as
given 
\begin{equation}
\left(f^{2}W^{\prime 2}r^{2}+(r^{2}m^{2}+\left( \partial _{2}H\right)
^{2})f-r^{2}(E-eA_{0})^{2}\right) ^{3}m^{2}=0.
\end{equation}

Solving this equation for the radial part leads to the following integral 
\begin{equation}
W_{\pm }=\pm \int dr\frac{1}{f}\sqrt{E^{2}-2EeA_{0}+e^{2}A_{0}^{2}-f\left(
m^{2}+\frac{\left( \partial _{2}H\right) ^{2}}{r^{2}}\right) }.
\end{equation}

Now, we can expand the function $f(r)$ in Taylor's series near the horizon 
\begin{equation}
f(r_{\theta +})\approx f^{\prime }(r_{\theta +})(r-r_{\theta +}),
\end{equation}%
and by integrating around the pole at the outer horizon $r_{\theta +}$,
gives \cite{ang,sri1} 
\begin{equation}
W_{\pm }=\pm \frac{i\pi \sqrt{E^{2}-2EeA_{0}+e^{2}A_{0}^{2}}}{f^{\prime
}|_{r_{\theta +}}}.  \label{integral}
\end{equation}

Since every outside particle falls into the black hole with a $100\%$ chance
of entering the black hole , therefore, the corresponding probability of the
ingoing massive vector particle must be set 
\begin{equation}
P_{-}\simeq e^{-2ImW_{-}}=1
\end{equation}%
which implies also that 
\begin{equation}
ImS_{-}=ImW_{-}+ImC=0.  \label{ingoing}
\end{equation}

On the other hand, for the outgoing particle we have 
\begin{equation}
ImS_{+}=ImW_{+}+ImC,
\end{equation}%
but we know that $ImC=-ImW_{-}$, and also from the Eq.\eqref{integral} we
see that 
\begin{equation}
W_{+}=-W_{-}.
\end{equation}

Therefore, the probability for the outgoing vector massive particle is given
by 
\begin{equation}
P_{+}=e^{-2ImS}\simeq e^{-4ImW_{+/-}}.
\end{equation}

Finally, the tunneling rate of massive vector particles tunneling from
inside to outside the horizon is given by 
\begin{equation}
\Gamma =\frac{P_{+}}{P_{-}}\simeq e^{(-4ImW_{+})}=e^{-\frac{E_{net}}{T_{H}}},
\label{prob}
\end{equation}

where $E_{net}=E-eA_{0}$. Then, the Hawking temperature of the
noncommutative Reissner-Nordstr\"{o}m black holes is recovered as 
\begin{equation}
T_{H}=\frac{1}{4\pi}\frac{df(r_{\theta+})}{dr}.
\end{equation}

Using Eq.\eqref{f}, one can recover the Hawking temperature for
noncommutative charged black hole, which is in agreement with \cite%
{nicolini2,nicolini5,hamid} 
\begin{equation}
T_{H}=\frac{1}{4\pi r_{\theta +}}\left[ 1-\frac{r_{\theta +}^{3}\exp {\left(
-\frac{r_{\theta +}^{2}}{4\theta +}\right) }}{4\theta ^{3/2}\gamma \left( 
\frac{3}{2},\frac{r_{\theta +}^{2}}{4\theta }\right) }\right] -\frac{Q^{2}}{%
\pi ^{2}r_{\theta +}^{3}}\left[ \gamma ^{2}\left( \frac{3}{2},\frac{%
r_{\theta +}^{2}}{4\theta }\right) +\frac{r_{\theta +}^{3}\exp {\left( -%
\frac{r_{\theta +}^{2}}{4\theta }\right) }\left( \gamma ^{2}\left( \frac{1}{2%
},\frac{r_{\theta +}^{2}}{4\theta }\right) -\frac{r_{\theta +}}{\sqrt{%
2\theta }}\gamma \left( \frac{1}{2},\frac{r_{\theta +}^{2}}{2\theta }\right)
\right) }{16\,\theta ^{3/2}\gamma \left( \frac{3}{2},\frac{r_{\theta +}^{2}}{%
4\theta }\right) }\right]   \label{htem}
\end{equation}

\bigskip 

\textit{Therefore, one can recovers the classical Hawking temperature $T_{H}=\frac{%
\sqrt{M^{2}-Q^{2}}}{2\pi \left( M+\sqrt{M^{2}-Q^{2}}\right) ^{2}}$ on the
commutative case. The Eq.(\ref{htem}) \ shows that in the noncommutative
framework, there is not completely evaporation of the black hole inasmuch as
it produces a Planck-sized remnant including the information. So information
might be preserved in this remnant. The simple $r_{\pm }$ dependent form of
the Hawking temperature from the noncommutative RN black hole can be
approximately written as follows $T_{H}=\frac{\kappa \left( M,Q\right) }{2\pi 
}\simeq \frac{1}{4\pi }\frac{r_{\theta +}-r_{\theta -}}{r_{\theta +}^{2}}$,
where the surface gravity \ $\kappa \left( M,Q\right) $ is $\frac{r_{\theta
+}-r_{\theta -}}{2r_{\theta +}^{2}}.$ }

\section{Tunneling From Charged Noncommutative BTZ Black Holes}

The line element describing the BTZ black hole in a noncommutative
background given by~\cite{chang,brito}: 
\begin{equation}
ds^{2}=-F^{2}dt^{2}+H^{-2}dr^{2}+2r^{2}N^{\phi }dtd\phi +\left( r^{2}-\frac{%
\theta B}{2}\right) d\phi ^{2},  \label{mbtz}
\end{equation}%
where the metric components are 
\begin{eqnarray}
&&F^{2}=\frac{r^{2}-r_{+}^{2}-r_{-}^{2}}{l^{2}}-\frac{\theta B}{2},
\label{F} \\
&&H^{2}=\frac{1}{r^{2}l^{2}}\left[ (r^{2}-r_{+}^{2})(r^{2}-r_{-}^{2})-\frac{%
\theta B}{2}(2r^{2}-r_{+}^{2}-r_{-}^{2})\right] ,  \label{N} \\
&&N^{\phi }=-\frac{r_{+}r_{-}}{l^{2}r^{2}}.
\end{eqnarray}%
where $B$ and $\theta $ are the magnitude of the magnetic field, and the
noncommutative parameter, respectively. Furthermore, $r_{+}$ and $r_{-}$ are
the outer and inner horizons. The event horizons are given as 
\begin{equation}
r_{\pm }^{2}=\frac{l^{2}M}{2}\left[ 1\pm \sqrt{1-\left( \frac{J}{Ml}\right)
^{2}}\right] .
\end{equation}%
The metric given in Eq.(\ref{mbtz}) can be rewritten as 
\begin{equation}
ds^{2}=-fdt^{2}+Q^{-1}dr^{2}+\frac{J}{r}rdrdt+\left( 1-\frac{\theta B}{2r^{2}%
}\right) r^{2}d\phi ^{2},  \label{mbtzz}
\end{equation}%
where 
\begin{eqnarray}
&&f=-M+\frac{r^{2}}{l^{2}}-\frac{\theta B}{2}, \\
&&Q=-M+\frac{r^{2}}{l^{2}}+\frac{J^{2}}{4r^{2}}-\frac{\theta B}{2}\left( 
\frac{2}{l^{2}}-\frac{M}{r^{2}}\right) .  \label{Q}
\end{eqnarray}%
Here M is the mass, J is the angular spin of the BTZ black hole. Noted that
the BTZ black hole without electric charge can be obtained as the quotient
of AdS space. The vector potential of the black hole is given by 
\begin{equation}
A_{\mu }=(A_{t},0,0),
\end{equation}

where $A_{t}=-Q\ln (\frac{r}{l})$.

To use the WKB approximation, as similar way done in section 2, we insert
the wave function given in Eq.(\ref{ans1}) and the vector potential into the
Proca equation of motion for the $W$-boson field \ of Eq.(\ref{proca2}) for
the near the event horizon of a noncommutative BTZ black hole in the
condition of $J=0$ and the metric is given as follows 
\begin{equation}
ds^{2}=-\tilde{f}dt^{2}+\tilde{Q}^{-1}dr^{2}+\left( r^{2}-K\right) d\phi
^{2},  \label{btzzmet}
\end{equation}%
where $K=\frac{\theta B}{2},$ $\tilde{f}=f^{\prime }(\hat{r}_{+})(r-\hat{r}%
_{+})$ and $\tilde{Q}=Q^{\prime }(\hat{r}_{+})(r-\hat{r}_{+})$. Dividing by
the exponential term and considering the leading terms yield three equations
;\bigskip 
\begin{eqnarray}
0 &=&\left[ \frac{Q(r^{2}-K)\left( \partial _{r}S\right) ^{2}+\left(
\partial _{\phi }S\right) ^{2}+m^{2}(r^{2}-K)}{f(r^{2}-K)}\right] C_{0}+%
\left[\frac{-Q\left[(\partial _{t}S)+A_{0}e\right] \partial _{r}S}{f}\right]%
C_{1}  \notag \\
&&+\left[\frac{-(\partial_{\phi }S)((\partial _{t}S)+eA_{0})}{f(r^{2}-K)}%
\right]C_{2},
\end{eqnarray}

\begin{equation}
0=\left[ -\frac{(\partial _{r}S)(\partial _{t}S+eA_{0})}{f}\right] C_{0}+%
\left[ \frac{(\partial _{\phi }S)(\partial _{r}S)}{r^{2}-K}\right] C_{2}+%
\left[ \frac{(r^{2}-K)(\partial _{t}S+eA_{0})^{2}-f(\partial _{\phi
}S)^{2}-m^{2}f(r^{2}-K)}{f(r^{2}-K)}\right] C_{1}
\end{equation}%
\begin{equation}
0=\frac{\left[ (\partial _{\phi }S)\left( (\partial _{t}S)+eA_{0}\right) %
\right] }{f}C_{0}+\left[ -Q(\partial _{\phi }S)(\partial _{r}S)\right] C_{1}+%
\left[ Q(\partial _{r}S)^{2}-\frac{((\partial _{t}S)+eA_{0})^{2}+m^{2}f}{f}%
\right] C_{2}.
\end{equation}

Using the spacetime symmetries one may choose the following ansatz 
\begin{equation}
S=-Et+W(r)+j\varphi +c,
\end{equation}%
which leads to the following non-zero elements of the coefficient matrix $%
\mathbb{Z}
$ are given by 
\begin{eqnarray}
\mathbb{Z}
_{11} &=&\left( \frac{QW^{\prime 2}+m^{2}}{f}\right) , \\
\mathbb{Z}
_{12} &=&-\frac{QW^{\prime }}{f}(eA_{0}-E) \\
\mathbb{Z}
_{21} &=&-\frac{W^{\prime }}{f}(eA_{0}-E) \\
\mathbb{Z}
_{22} &=&\frac{(eA_{0}-E)^{2}-m^{2}f}{f} \\
\mathbb{Z}
_{33} &=&\frac{QfW^{\prime 2}-(eA_{0}-E)^{2}+m^{2}f}{f}.
\end{eqnarray}

Using the space-time symmetries of the metric Eq.(\ref{btzzmet}), we can
and the nontrivial solution of this equation 
\begin{equation}
\mathbb{Z}
(C_{0},C_{1},C_{2})^{T}=0,
\end{equation}

is obtained by using the the determinant of the matrix $%
\mathbb{Z}
$ equals zero as given 
\begin{equation}
m^{2}\left(e^{2}A_{0}^{2}-2eEA_{0}+E^{2}-f(m^{2}+QfW^{\prime
2})\right)^{2}=0.
\end{equation}

Solving this equation for the radial part leads to the following integral 
\begin{equation}
W_{\pm }=\pm \int dr\frac{1}{\sqrt{Qf}}\sqrt{%
E^{2}-2EeA_{0}+e^{2}A_{0}^{2}-m^{2}f}.
\end{equation}

Solving this integral by integrating around the pole, gives 
\begin{equation}
W_{\pm }=\pm \frac{i\pi \sqrt{E^{2}-2EeA_{0}+e^{2}A_{0}^{2}}}{\sqrt{%
Q^{\prime }f^{\prime }}|_{\hat{r}_{+}}}.
\end{equation}

By using similar arguments as in the last section, for the tunneling rate we
must have 
\begin{equation}
\Gamma \simeq e^{(-4ImW_{+})}=e^{-\frac{E_{net}}{T_{H}}},
\end{equation}

where $E_{net}=E-eA_{0}$.

In this way, Hawking temperature at the event horizon of the charged
noncommutative BTZ black holes is calculated as 
\begin{equation}
T_{H}=\frac{1}{4\pi }\sqrt{f^{\prime }(\hat{r}_{+})Q^{\prime }(\hat{r}_{+})}=%
\frac{\hat{r}_{+}}{2\pi l^{2}}\left( 1-\frac{\theta Br_{+}^{2}}{4\hat{r}%
_{+}^{4}}\right) +\hdots.
\end{equation}

Finally, the last result in terms of $r_{+}$, gives 
\begin{equation}
T_{H}=\frac{r_{+}}{2\pi l^{2}}\left(1-\frac{\theta^{2} B^{2}}{16r_{+}^{4}}%
\right)+\hdots
\end{equation}

Thus, we have recovered the Hawking temperature with remnant for the charged
noncommutative BTZ black hole \cite{chang}. As expected, there are
corrections for the Hawking temperature due to the noncommutativity of
spacetime. 

\section{GUP-Entropy and Corrected Hawking Temperature of the Noncommutative 
\textit{RN} Black Holes}

In this section, by using the recent works in (\cite{brit,gup1,gup2}), we
show the effect of the GUP to the Noncommutative \textit{RN }
thermodynamics. Firstly, the GUP expression with a quadratic term in
momentum is written as \cite{gup3} 
\begin{equation}
\Delta x\Delta p\geq \hbar \left( 1-\beta \Delta p+\beta ^{2}(\Delta
p)^{2}\right) ,
\end{equation}%
where the dimensionless constant $\beta $ ( $\beta =\alpha l_{p}/\hbar )$,
here $\alpha $ is a positive constant, $l_{p}$ and $\hbar $, are the Planck
length ($l_{p}=\sqrt{\hbar G/c^{3}}\approx 10^{-35}m)$ and the Planck
constant respectively. After some algebraic manipulation the last equation
reduces to

\begin{equation}
\Delta p\geq \frac{\hbar (\Delta x+\alpha l_{p})}{2\alpha ^{2}l_{p}^{2}}%
\left( 1-\sqrt{1-\frac{4\alpha ^{2}l_{p}^{2}}{(\Delta x+\alpha l_{p})^{2}}}%
\right).  \label{deltapp}
\end{equation}

Note that $l_{p}/\Delta x$ is infinitesimally small compared with unity, one
can therefore expand the last equation in Taylor series as follows 
\begin{equation}
\Delta p\geq \frac{1}{2 \Delta x}\left[ 1-\frac{\alpha }{2\Delta x}+\frac{%
\alpha ^{2}}{2(\Delta x)^{2}}+\cdots \right] .  \label{p}
\end{equation}%
It is noted that after using the constant as a one ($G=c=k_{B}=1$, $\hbar =1$%
, and $l_{p}=1)$, uncertainty principle reduces to $\Delta x\Delta p\geq 1.$
The saturated form of the Heisenberg uncertainty principle depended on
energy of a particle can be also written \ in the form of $E\Delta x\geq 1.$
Consequently, the new form of the Eq.(\ref{deltapp}) is 
\begin{equation}
E_{GUP}\geq E\left[ 1-\frac{\alpha }{2(\Delta x)}+\frac{\alpha ^{2}}{%
2(\Delta x)^{2}}+\cdots \right] ,
\end{equation}%
where $E_{GUP}$ is the corrected energy of the particle. Now we can rewrite
the tunneling probability which we find in Eq.(\ref{prob}), by considering
the effect of GUP as follows 
\begin{equation}
\Gamma \simeq \exp [-2\mathrm{{\,Im}{(\mathcal{S}})]=\exp \left[ \frac{-2{%
\pi }E_{GUP}}{\kappa }\right] },
\end{equation}%
where $\kappa $ is the surface gravity of noncommutative \textit{RN} black
hole horizon given by $\kappa =2f^{\prime }(r_{\theta _{+}})$. Comparing
this with the Boltzmann factor the GUP-corrected black hole temperature of
noncommutative \textit{RN} is found as 
\begin{equation}
T_{GUP}=T_{H}\left[ 1-\frac{\alpha }{2(\Delta x)}+\frac{\alpha ^{2}}{%
2(\Delta x)^{2}}+\cdots \right] ^{-1},
\end{equation}%
where $\Delta x=2r_{\theta _{+}}=2\left( M_{\theta _{+}}+\sqrt{M_{_{\theta
_{+}}}^{2}-Q_{_{\theta _{+}}}^{2}}\right) .$ Thus, the corrected Hawking
temperature due to the GUP-effects, reads 
\begin{equation}
T_{GUP}=\left( \frac{1}{4\pi r_{\theta +}}\left[ A\right] -\frac{4Q^{2}}{\pi
r_{\theta +}^{3}}\left[ B\right] \right) \times \left( 1+\frac{\alpha }{%
4r_{\theta _{+}}}-\frac{\alpha ^{2}}{8r_{\theta _{+}}^{2}}+\cdots \right) 
\end{equation}

where 
\begin{equation}
A=1-\left[ \frac{r_{\theta +}^{3}\exp {\left( -\frac{r_{\theta +}^{2}}{%
4\theta +}\right) }}{4\theta ^{3/2}\gamma \left( \frac{3}{2},\frac{r_{\theta
+}^{2}}{4\theta }\right) }\right] ,
\end{equation}%
and 
\begin{equation}
B=\gamma ^{2}\left( \frac{3}{2},\frac{r_{\theta +}^{2}}{4\theta }\right) +%
\frac{r_{\theta +}^{3}\exp {\left( -\frac{r_{\theta +}^{2}}{4\theta }\right) 
}\left( \gamma ^{2}\left( \frac{1}{2},\frac{r_{\theta +}^{2}}{4\theta }%
\right) -\frac{r_{\theta +}}{\sqrt{2\theta }}\gamma \left( \frac{1}{2},\frac{%
r_{\theta +}^{2}}{2\theta }\right) \right) }{16\,\theta ^{3/2}\gamma \left( 
\frac{3}{2},\frac{r_{\theta +}^{2}}{4\theta }\right) }.
\end{equation}

It is noted that for $\alpha =0$, it is obtained the noncommutative \textit{%
RN} black hole temperature in Eq.\ (\ref{htem}). Therefore, we get more
remnants for the noncommutative \textit{RN} black hole. Now let us compute
the GUP-entropy corrections, for this reason we can now use the first law of
BH thermodynamics and let for simplicity choose $\Delta x=2r_{\theta _{+}}$.
It follows 
\begin{eqnarray}
S_{GUP} &=&\int \frac{{dE}}{T_{GUP}}=\int \frac{\kappa \,dA_{h}}{8\pi T_{GUP}%
}  \notag \\
&=&\int 2\pi r_{\theta _{+}}dr_{\theta _{+}}\left[ 1-\frac{\alpha }{%
4r_{\theta _{+}}}+\frac{\alpha ^{2}}{8r_{\theta _{+}}^{2}}+\cdots \right] , 
\notag \\
&=&\pi r_{\theta _{+}}^{2}-\frac{\alpha \pi }{2}r_{\theta _{+}}+\frac{\alpha
^{2}\pi }{4}\ln r_{\theta _{+}}+\cdots 
\end{eqnarray}%
in the first and second line of the last equation we have used $\kappa =2\pi
T_{H}$ and $dA_{h}=8\pi r_{\theta _{+}}dr_{\theta _{+}}$, respectively.
Finally, we can rewrite the last equation in terms of the surface area of
the event horizon as follows 
\begin{equation}
S_{GUP}=\frac{A_{h}}{4}-\frac{\alpha }{4}\sqrt{A_{h}\pi }+\frac{\pi \alpha
^{2}}{8}\ln \frac{A_{h}}{4}+\cdots 
\end{equation}

Notice that for $\alpha _{_{GUP}}=0$, we recover the well-known area law for
the BH mechanics i.e., $\left. S_{GUP}\right\vert _{\alpha =0}\rightarrow
S=A_{h}/4$. Remnants mainly due to extraordinary large amount of entropy
confined within a tiny volume.

Note that in order to write explicitly the entropy in terms of $Q$, $M$ and
the parameter $\theta $, we need to recall the following relation: 
\begin{equation}
r_{\theta _{+}}=M_{\theta _{+}}+\sqrt{M_{\theta _{+}}^{2}-Q_{\theta _{+}}^{2}%
},
\end{equation}%
where 
\begin{equation}
M_{\theta _{+}}=M\left[ \epsilon \left( \frac{M+\sqrt{M^{2}-Q^{2}}}{2\sqrt{%
\theta }}\right) -\left( \frac{M+\sqrt{M^{2}-Q^{2}}}{\sqrt{\pi \theta }}%
\right) \exp \left( -\frac{(M+\sqrt{M^{2}-Q^{2}})^{2}}{4\theta }\right) %
\right] ,
\end{equation}%
\begin{equation}
Q_{\theta _{+}}=Q\sqrt{\epsilon ^{2}\left( \frac{M+\sqrt{M^{2}-Q^{2}}}{2%
\sqrt{\theta }}\right) -\frac{\left( M+\sqrt{M^{2}-Q^{2}}\right) ^{2}}{\sqrt{%
2\pi \theta }}\exp \left( -\frac{(M+\sqrt{M^{2}-Q^{2}})^{2}}{4\theta }%
\right) },
\end{equation}%
where $\epsilon (x)$ is the Gauss error function defined as 
\begin{equation}
\epsilon (x)=\frac{2}{\sqrt{\pi }}\int_{0}^{x}e^{-u^{2}}du.
\end{equation}

As expected, for very large masses the function $\epsilon(x)$ tends to unity
while the exponential term goes to zero, giving the classical \textit{RN}
horizons $r_{\theta_{+}}\to r_{+}=M+\sqrt{M^{2}-Q^{2}}$.

\section{Conclusion}

In summary, we have investigated the tunneling of charged massive bosons $%
W^{\pm }$ from charged noncommutative RN black holes and BTZ black holes.
The corresponding tunneling rates and the expected Hawking temperatures have
been recovered by using the WKB approximation and the Hamilton-Jacobi
equation in both cases. Furthermore, the results in this paper are
consistent with an underlying unitary theory.

\textit{Last but not least, it is also shown that the remnants of the
hawking temperature from the charged noncummutative black holes are
increased by the effect of GUP. Therefore, this GUP-corrected Hawking
radiation is common use for entire black holes. Then we have derived the
logarithmic corrections in the leading order for the corrected entropy.\
Noted that the existence of the remnant depends critically on the full form
of the entropy corrected by GUP. The corrected Hawking temperature and
entropy are shown to depend on the noncommutaive parameter $\theta $ and
quantum gravity effects $\alpha $. \ The black hole remnant would offer a
way in which information might be preserved. Hence, the existence of a
remnant may have important experimental consequences for the detection of
black holes at the the Large Hadron Collider (LHC).}
\section{Acknowledgment}
The authors would like to thank the anonymous reviewers.


\begin{thebibliography}{99}
\bibitem{Hawking1} S. W. Hawking, Commun. Math. Phys. \textbf{43}, 199
(1975); erratum-ibid, \textbf{46}, 206 (1976).

\bibitem{gibbons} G. W. Gibbons and S. W. Hawking, Phys. Rev. D \textbf{15},
2738 (1977).

\bibitem{sh1} I. Sakalli, M. Halilsoy, and H. Pasaoglu, Astrophys.Space Sci. 
\textbf{340}, 155 (2012).

\bibitem{kraus1} P. Kraus, F. Wilczek, Mod. Phys. Lett. A \textbf{9}, 3713
(1994).

\bibitem{kraus2} P. Kraus, F. Wilczek, Nucl. Phys. B \textbf{437}, 231
(1995).

\bibitem{perkih1} M.K. Parikh, F. Wilczek, Phys. Rev. Lett. \textbf{85},
5042 (2000).

\bibitem{perkih2} M.K. Parikh, Phys. Lett. B \textbf{546}, 189 (2002).

\bibitem{perkih3} M.K. Parikh, Int. J. Mod. Phys. D \textbf{13}, 2351 (2004).

\bibitem{ang} M. Angheben, M. Nadalini, L. Vanzo, S. Zerbini, J. High Energy
Phys. \textbf{05}, 014 (2005).

\bibitem{sri1} K. Srinivasan, T. Padmanabhan, Phys. Rev. D \textbf{60},
024007 (1999).

\bibitem{sri2} S. Shankaranarayanan, K. Srinivasan, T. Padmanabhan, Mod.
Phys. Letts. \textbf{16}, 571 (2001)

\bibitem{ahmed} J. Ahmed, K. Saifullah, JCAP \textbf{08}, 011 (2011).

\bibitem{kimet} K. Jusufi, Astrophys. Space Sci. \textbf{360}, 22 (2015).

\bibitem{vanzo} L. Vanzo, G. Acquaviva, R. Di Criscienzo, Class. Quantum
Gravity \textbf{28}, 18 (2011).

\bibitem{mann0} R. Kerner, R.B. Mann, Phys. Rev. D \textbf{73}, 104010
(2006).

\bibitem{mann1} R. Kerner and R.B. Mann, Class. Quant. Grav. \textbf{25},
095014 (2008).

\bibitem{mann2} R. Kerner and R.B. Mann, Phys. Lett. B \textbf{665}, 277-283
(2008).

\bibitem{sh2} I. Sakalli, A. Ovgun, Astrophys Space Sci. \textbf{359}, 32
(2015).

\bibitem{mann3} A. Yale, R. B. Mann Phys. Lett. B \textbf{673}, 168-172,
(2009).

\bibitem{kruglov1} S.I Kruglov, Mod. Phys. Lett. A \textbf{29}, 1450203
(2014).

\bibitem{kruglov2} S.I Kruglov, Int. J. Mod. Phys. A \textbf{29}, 1450118
(2014).

\bibitem{sh11} I. Sakalli, A. Ovgun, J.Exp.Theor.Phys, \textbf{121}, 3,
404-407, (2015).

\bibitem{sh3} I. Sakalli, A. \"{O}vg\"{u}n, Gen.Rel.Grav. \textbf{48,} 1
(2016).

\bibitem{sh4} I. Sakalli, A. Ovgun, Eur. Phys. J. Plus \textbf{130}: 110,
(2015).

\bibitem{huang} G. Chen, Y. Huang, Int. J. Mod. Phys. A \textbf{30}, 15,
(2015).

\bibitem{chenn} G. Chen, S. Zhou, Y. Huang, Astrophys Space Sci \textbf{357}%
:51, (2015).

\bibitem{ao} A. \"{O}vg\"{u}n, International Journal of Theoretical Physics,
1-9, (2016) , arXiv:1508.04100.

\bibitem{nicolini1} P. Nicolini, A. Smailagic, E. Spallucci, Phys. Lett. B 
\textbf{632}, 547, (2006).

\bibitem{nicolini2} S. Ansoldi, P. Nicolini, A. Smailagic, E. Spallucci,
Phys. Lett. B \textbf{645}, 261-266, (2007).

\bibitem{nicolini3} E. Spallucci, A. Smailagic, P. Nicolini, Phys. Lett. B 
\textbf{670} 449-454, (2009).

\bibitem{nicolini4} L. Modesto, P. Nicolini, Phys. Rev. D \textbf{82},
104035, (2010).

\bibitem{sma1} A. Smailagic and E. Spallucci, J. Phys. A \textbf{36}, L467
(2003).

\bibitem{sma2} A. Smailagic and E. Spallucci, J. Phys. A \textbf{36}, L517
(2003).

\bibitem{sma3} A. Smailagic and E. Spallucci, J. Phys. A \textbf{37}, 7169
(2004).

\bibitem{hamid} S. Hamid Mehdipour, Int. J. Mod. Phys. A \textbf{25},
5543-5555, (2010).

\bibitem{zhao} Y. Miao, Z. Xue, S. Zhang, Gen. Relativ. Gravit. \textbf{44}
555-566, (2012).

\bibitem{nozari1} K. Nozari, S. H. Mehdipour, JHEP 0903, \textbf{061},
(2009).

\bibitem{nozari2} K. Nozari, S. H. Mehdipour, Class. Quant. Grav. \textbf{25}%
, 175015, (2008).

\bibitem{gup1} M. Maggiore, Phys.Lett. B\textbf{\ 304}, 65-69 (1993)  

\bibitem{gup2} R. J. Adler, P. Chen, D. I. Santiago, Gen.Rel.Grav. \textbf{33%
}, 2101-2108 (2001).

\bibitem{gup3} A. Kempf, G. Mangano, R. B. Mann, Phys.Rev. D \textbf{52},
1108-1118 (1995). 

\bibitem{gup4} M. Faizal, M. M. Khalil, Int.J.Mod.Phys. A \textbf{30},
no.22, 1550144, (2015).

\bibitem{brit} M.A. Anacleto, F.A. Brito, A.G. Cavalcanti, E. Passos, J.
Spinelly, arXiv:1510.08444.

\bibitem{xiang} X. Li, G. Chen, Physics Letters B \textbf{751} 34-38, (2015)
.

\bibitem{nicolini5} P. Nicolini, Int. J. Mod. Phys. A \textbf{24},
1229-1308, (2009).

\bibitem{chang} H. Kim, M. Park, C. Rim, J. H. Yee, JHEP \textbf{10 , }060,
(2008) ; E.Chang -Young, D. Lee, Y. Lee, Class. Quant. Grav. \textbf{26},
185001, (2009).

\bibitem{btz} M. Banados, C. Teitelboim, J. Zanelli, Phys.Rev.Lett. \textbf{%
69}, 1849-1851 (1992).

\bibitem{btz1} M. Ammon, M. Gutperle, P. Kraus, E. Perlmutter, J.Phys. A 
\textbf{46}, 214001 (2013). 

\bibitem{brito} M. A. Anacleto, F. A. Brito, E. Passos, Phy. Lett. B \textbf{%
743}, 184--188, (2015).

\bibitem{jafer} J. Sadeghi and V.R. Shajiee, Int. J. Theor. Phys. \textbf{07}%
, July (2015).

\bibitem{Hendi} S. H. Hendi, S. Panahiyan and R. Mamasani, Gen. Rel. Grav. 
\textbf{47}, no. 8, 91 (2015).

\bibitem{gup1} M. A. Anacleto, F . A. Brito, and E. Passos, Phys. Lett. B 
\textbf{749}, 181 (2015).

\bibitem{gup2} M. A. Anacleto, F . A. Brito, G . C. Luna, E. Passos, and J.
Spinelly, Ann. Phys. \textbf{362}, 436 (2015).

\bibitem{gup3} A. F. Ali, S. Das, and E. C. Vagenas, Phys. Lett. B \textbf{%
678}, 497 (2009).

\bibitem{kahn} Y. Kahn and J. Thaler, Phys. Rev. D \textbf{86}, 115012
(2012).

\bibitem{bednyakov} V. A. Bednyakov, arXiv:1505.04380.

\bibitem{balewski} J. Balewski, J. Bernauer, J. Bessuille, R. Corliss , R.
Cowan, et al., arXiv:1412.4717.

\bibitem{emidio} E. Gabrielli, L. Marzola, M. Raidal and H. Veermae, JHEP 
\textbf{1508}, 150 (2015).
\end{thebibliography}
\end{document}